\documentclass{article}
\usepackage[utf8]{inputenc}
\usepackage{authblk}
\usepackage{parskip}
\usepackage{bm}
\usepackage{amsmath,amssymb,amsthm,amsfonts}
\usepackage[style=nature, citestyle=numeric-comp, backend=biber, sorting=none, bibencoding=utf8]{biblatex}
\usepackage{graphicx}
\usepackage[labelfont=bf]{caption}
\usepackage{todonotes}
\usepackage{algorithm}
\usepackage{algpseudocode}
\newcommand\numberthis{\addtocounter{equation}{1}\tag{\theequation}}
\makeatletter
\renewcommand{\ALG@name}{Supplementary Algorithm}
\makeatother
\newcommand{\beginsupplement}{%
        \setcounter{table}{0}
        \renewcommand{\thetable}{S\arabic{table}}%
        \setcounter{figure}{0}
        \renewcommand{\thefigure}{S\arabic{figure}}%
        \setcounter{section}{0}
        \renewcommand{\thesection}{SI\arabic{section}}%
        \setcounter{equation}{0}
        \renewcommand{\theequation}{S\arabic{equation}}
     }
\makeatletter
\newenvironment{breakablealgorithm}
  {
   \begin{center}
     \refstepcounter{algorithm}
     \hrule height.8pt depth0pt \kern2pt
     \renewcommand{\caption}[2][\relax]{
       {\raggedright\textbf{\fname@algorithm~\thealgorithm} ##2\par}%
       \ifx\relax##1\relax 
         \addcontentsline{loa}{algorithm}{\protect\numberline{\thealgorithm}##2}%
       \else 
         \addcontentsline{loa}{algorithm}{\protect\numberline{\thealgorithm}##1}%
       \fi
       \kern2pt\hrule\kern2pt
     }
  }{
     \kern2pt\hrule\relax
   \end{center}
  }
\makeatother
\usepackage[margin=1.0in]{geometry}
\usepackage[para]{footmisc}
\title{Confidence and second-order errors in cortical circuits}

\author{Arno Granier\thanks{corresponding author: arno.granier@unibe.ch}}
\author{Mihai A. Petrovici}
\author{Walter Senn\thanks{These authors jointly supervised this work.}}
\author{Katharina A. Wilmes$^\dagger$}
\affil[1]{Department of Physiology, University of Bern, Switzerland}

\begin{document}
\setlength{\abovedisplayskip}{5pt}
\setlength{\belowdisplayskip}{5pt}

\maketitle

\begin{abstract}\noindent Minimization of cortical prediction errors has been considered a key computational goal of the cerebral cortex underlying perception, action and learning.
However, it is still unclear how the cortex should form and use information about uncertainty in this process.
Here, we formally derive neural dynamics that minimize prediction errors under the assumption that cortical areas must not only predict the activity in other areas and sensory streams but also jointly project their confidence (inverse expected uncertainty) in their predictions.
In the resulting neuronal dynamics, the integration of bottom-up and top-down cortical streams is dynamically modulated based on confidence in accordance with the Bayesian principle. 
Moreover, the theory predicts the existence of cortical second-order errors, comparing confidence and actual performance. These errors are propagated through the cortical hierarchy alongside classical prediction errors and are used to learn the weights of synapses responsible for formulating confidence.
We propose a detailed mapping of the theory to cortical circuitry, discuss entailed functional interpretations and provide potential directions for experimental work. \end{abstract}

\section{Introduction}

Taking uncertainty into account in models of cortical processing has proven beneficial to capture behavioral and neural data at multiple scales \cite{pouget2013probabilistic, koblinger2021representations, walker2023studying}. Empirical studies on humans and other animals show that prior knowledge and data from multiple modalities are weighted by their relative uncertainty during perceptual integration \cite{ernst2002humans}, decision-making \cite{behrens2007learning, kiani2009representation} and sensorimotor control \cite{kording2004bayesian, darlington2018neural}. Crucially, uncertainty is context-dependent and can vary dynamically \cite{fetsch2009dynamic, noppeney2021perceptual}. For example, in the dark, animals should rely more on prior knowledge of the environment than vision, whereas in daylight, they can trust their vision more.

Additionally, cortical processing has been described based on the notion of prediction \cite{de2018expectations,teufel2020forms}, with cortical areas attempting to predict the activity in other areas or sensory streams.
The computational goal of the cortex would then be to minimize differences between these predictions and actual activity, commonly referred to as prediction errors.
Neural computations realizing this goal have been proposed as canonical cortical computations \cite{rao1999predictive, friston2005theory, keller2018predictive}.
One way to incorporate uncertainty in these models is to assume that a cortical prediction should not simply be a single potential representation of the target area but rather a distribution over the space of possible representations.
In that case, normative theories based on variants of maximum likelihood estimation suggest that cortical prediction errors should be multiplicatively weighted by the inverse variance of the predictive distribution.
This modulatory weighting of prediction errors has gained a central place in the branch of cognitive sciences based on predictive coding \cite{friston2018does, yon2021precision}, most notably in models of attention \cite{feldman2010attention, kok2012attention, jiang2013attention} and in neuropsychiatry \cite{van2014precise, sterzer2018predictive, corlett2019hallucinations, friston2022computational}.
Potential implementations in the cerebral cortex have been discussed, notably in cortico-pulvinar loops \cite{kanai2015cerebral} or more generally through neuromodulation \cite{angela2005uncertainty, lawson2021computational}. 
However, a formal account of the role of learned and context-dependent uncertainty estimation is still missing. 

In this work, we suppose that cortical areas must not only predict the activity in other areas and sensory streams but also jointly estimate the confidence of their predictions, where we define confidence as the (estimated) inverse expected uncertainty of the prediction. In other words, we introduce measures of confidence computed at each level of the cortical hierarchy as a function of current higher-level representations, forming a hierarchy of confidence analogous to the hierarchy of predictions. For example, the representation of the environment will determine the degree of confidence in a prediction about the presence of a particular object (e.g.~this confidence will be different in familiar versus unfamiliar environments). Similarly, the representation of an object will determine the degree of confidence in predictions of lower-level features of that object (e.g.~some objects are always of the same color while others can vary).
This formulation is in line with rare instances where inverse uncertainty has been formulated as a function of current neuronal activity \cite{feldman2010attention, kanai2015cerebral}, and to be contrasted with the majority of literature in which it is predominantly defined as a parameter of the internal model, independent of current neuronal activity. With our formulation, confidence has a fast, dynamic, and context-dependent influence on neural dynamics, while the parameters of the function computing confidence, encoded in synaptic weights, slowly learn the statistics of the environment. 

\section{Results}

\subsection{An energy for cortical function}

Given the organization of the cortex into specialized areas, we define latent cortical representations as $\bm{u_1}, \dots, \bm{u_n}$, corresponding to the membrane potentials of neuronal populations in $n$ areas, and denote $\bm{u_0}$ the observation. 
For example, the observation $\bm{u_0}$ might be the activity of visual sensors (retina), and latent cortical representations $\bm{u_1}, \dots, \bm{u_n}$ might encode local orientation (V1), color (V4), objects (IT), etc.

\begin{figure}[!ht]%
        \centering
        \includegraphics{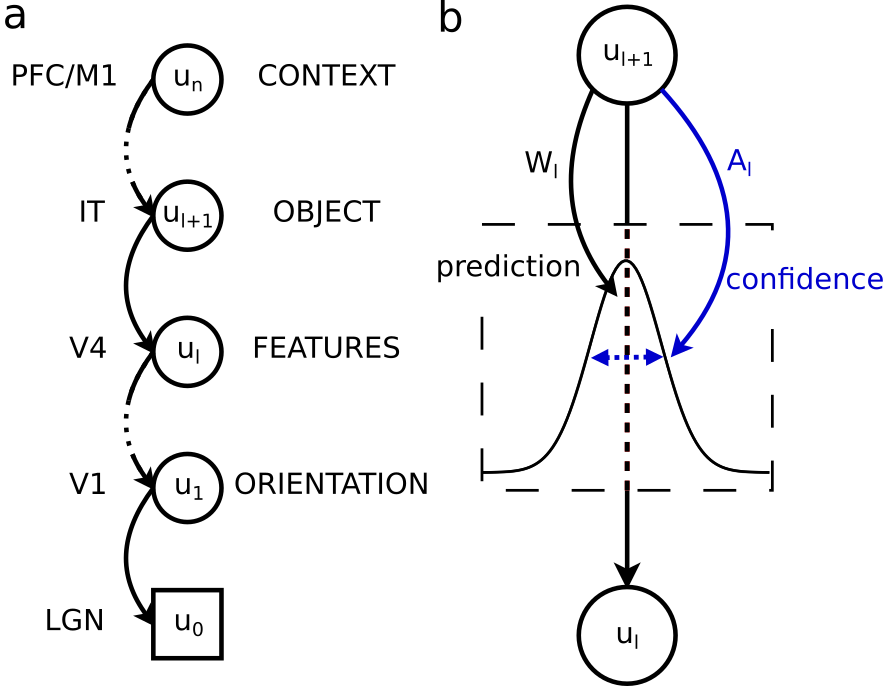}
        \caption{Predictive distributions in the cortical hierarchy. 
        (a) Probabilistic model. Latent representations ($\bm{u_{\ell}}$) are organized in a strict generative hierarchy. 
        (b) Predictions are Gaussian distributions. Both the mean ($\bm{\mu_{\ell}} = \bm{W_{\ell}r_{\ell+1}}$, first-order) and the confidence ($\bm{\pi_{\ell}} = \bm{A_{\ell}r_{\ell+1}}$, inverse variance, second-order) are functions of higher-level activity.}\label{fig:energy}
        \end{figure}

As a simplifying assumption, we organize areas in a strict generative hierarchy, such that area $l\!+\!1$ tries to predict the activity of only the area $l$ below (see Fig.~\ref{fig:energy}a).
It does so by sending its output rates $\bm{r_{\ell+1}} = \phi(\bm{u_{\ell+1}})$ through top-down synapses with plastic weights $\bm{W_{\ell}}$, where $\phi$ represents the neuronal activation function. 
Additionally, area $l\!+\!1$ similarly estimates and conveys to area $l$ the confidence of its prediction through top-down synapses with plastic weights $\bm{A_{\ell}}$. We further hypothesise that the resulting predictive distribution is the (entropy-maximizing) normal distribution with mean vector $\bm{\mu_{\ell}} = \bm{W_{\ell}}\bm{r_{\ell+1}}$ and confidence (inverse variance) vector $\bm{\pi_{\ell}} = \bm{A_{\ell}}\bm{r_{\ell+1}}$ (see Fig.~\ref{fig:energy}b). 
Crucially, confidence is not simply a static parameter of the model; instead, it is a parameterized function of current higher-level representations. For example, different context representations might lead to different levels of certainty about the presence of the same object, and different object representations might send more confident predictions for one sensory modality than another.
In essence, this is an extension of the notion of prediction, where cortical areas predict the confidence (second-order information) in addition to the mean (first-order information).

We can now formulate our energy (or cost) for cortical function 
\begin{equation}\label{eq:E}
E = \frac12\sum_{\ell=0}^{n-1} \|\bm{e_{\ell}}\|_{\bm{\pi_{\ell}}}^2 - \frac12\sum_{\ell=0}^{n-1} \log |\bm{\pi_{\ell}}|\;,
\end{equation}
where $\bm{e_{\ell}} = \bm{u_{\ell}} - \bm{\mu_{\ell}}$ is a prediction error, $\|\cdot\|_{\bm{\pi_{\ell}}}$ denotes the norm with $\bm{\pi_{\ell}}$ as a metric (i.e., a variance-normalized norm, $\|\bm{e_{\ell}}\|_{\bm{\pi_{\ell}}}^2=\bm{e_{\ell}}^T\text{diag}(\bm{\pi_{\ell}})\bm{e_{\ell}}$) and $|\cdot|$ denotes the product of components. This energy can be derived as the negative log-joint of a hierarchical generative probabilistic model (see Methods).
Note that $\|\bm{e_{\ell}}\|_{\bm{\pi_{\ell}}}$ is the classical Euclidean norm of standardized errors. In other words, here, we measure distances in terms of numbers of standard deviations away from the mean. This metric, the Mahalonobis distance, is a better measure of distance between a point (representation) and a Gaussian distribution (prediction) than simply the Euclidean distance to the mean $\|\bm{e_{\ell}}\|$.

This energy $E$ seems worth minimizing.
The first term is a measure of distance between actual representations and predictions, additionally taking into account the confidence of predictions: the more a prediction is confident, the more a deviation from it matters.
The second term indicates that high confidence is preferable. In other words, the cortex tries to reduce its expected uncertainty. That is, as long as high confidence does not excessively lead to an increase in the first term: there must be a balance between the confidence and the (average) magnitude of prediction errors. 
In other words, areas learn to be confident in predictions leading to small remaining errors ($\bm{e_{\ell-1}}^2$). Moreover, the second term also acts as a regularizer to avoid uninformative, i.e.\ very small, confidence.

Having formulated an energy for cortical function, we formally derive gradient-based neuronal dynamics and synaptic learning rules minimizing this energy.

\subsection{Neuronal dynamics with confidence estimation}
We classically derive neuronal dynamics of inference minimizing the energy $E$ through gradient descent. Moreover, we make use of confidence $\bm{\pi_{\ell}}$ as a metric to guide our descent \cite{surace2020choice}.
The resulting dynamics can be interpreted as an approximate second-order optimization scheme (see Methods).
This leads to the leaky neuronal dynamics
\begin{equation}\label{eq:udot}
    \tau\dot{\bm{u_{\ell}}} = - \bm{\pi_{\ell}}^{-1} \circ \partial E/\partial \bm{u_{\ell}} = -\bm{u_{\ell}} + \bm{\mu_{\ell}} + \bm{\pi_{\ell}}^{-1} \circ \bm{a_{\ell}}\;,
\end{equation}
integrating top-down predictions $\bm{\mu_{\ell}} = \bm{W_{\ell}}\bm{r_{\ell+1}}$, and total propagated errors
\begin{equation}\label{eq:a}
    \bm{a_{\ell}} = \bm{r'_{l}} \circ (\bm{W_{\ell-1}}^T(\bm{\pi_{\ell-1}} \circ \bm{e_{\ell-1}}) +  \bm{A_{\ell-1}}^T\bm{\delta_{\ell-1}})
\end{equation} 
defined as the sum of confidence-weighted prediction errors $\bm{\pi_{\ell-1}}\circ\bm{e_{\ell-1}}$ and second-order errors $\bm{\delta_{\ell-1}} = (\bm{\pi_{\ell-1}}^{-1} - \bm{e_{\ell-1}}^2)/2$, both propagated upwards from the lower area.
Here $\circ$ is the componentwise (Hadamard) product and $\bm{e_{\ell-1}}^2 = \bm{e_{\ell-1}}\circ\bm{e_{\ell-1}}$.
The second-order errors $\bm{\delta_{\ell}}$ are not errors on the prediction (of the mean) $\bm{\mu_{\ell}}$ but errors on the confidence $\bm{\pi_{\ell}}$, which are expected to be on average $0$ if and only if the estimate $\bm{\pi_{\ell}}$ correctly captures the underlying inverse variance.
Following previous work \cite{sacramento2018dendritic}, we suppose that total propagated errors $\bm{a_{\ell}}$ are encoded in the apical dendrites of cortical neurons with somatic membrane potential $\bm{u_{\ell}}$. 

These neuronal dynamics (Eqs.~\ref{eq:udot} and \ref{eq:a}) entail two major points of interest, one of gain modulation of errors based on confidence (see Fig.~\ref{fig:balancing}) and one of second-order error propagation (see Fig.~\ref{fig:deltaprop}).
In the following section, we complete our theoretical framework by deriving synaptic learning rules for parameters $\bm{W_{\ell}}$ and $\bm{A_{\ell}}$. We then return to neuronal dynamics and further unpack these two points of interest.

\subsection{Error-correcting synaptic learning of confidence}

At the equilibrium of neuronal dynamics, weights of synapses carrying predictions can be learned following the gradient
\begin{equation}\label{eq:Wdot}
    \dot{\bm{W_{\ell}}} \propto - \partial E/\bm{W_{\ell}} = (\bm{\pi_{\ell}}\circ\bm{e_{\ell}})\bm{r_{\ell+1}}^T \;,
\end{equation}
where $\bm{\pi_{\ell}}\circ\bm{e_{\ell}}$ are postsynaptic confidence-weighted prediction errors and $\bm{r_{\ell+1}}$ are presynaptic rates. This is the classical learning rule for prediction weights in the predictive coding framework\cite{rao1999predictive}.
By following this learning rule, synapses learn to correctly predict lower-level features (e.g.~orientation) from higher-level activity (e.g.~object). Additionally, confidence impacts learning speed: if a prediction is confident but wrong, a significant update is required, whereas an error on a prediction made with low confidence might reflect intrinsic variability and does not require a big update.

Similarly, weights $\bm{A_{\ell}}$ of synapses carrying confidence can also be learned following the gradient 
\begin{equation}\label{eq:Adot}
    \dot{\bm{A_{\ell}}} \propto - \partial E/\partial \bm{A_{\ell}} = \bm{\delta_{\ell}}\bm{r_{\ell+1}}^T\;,  
\end{equation}
where again $\bm{\delta_{\ell}}=(\bm{\pi_{\ell}}^{-1}-\bm{e_{\ell}}^2)/2$ are postsynaptic second-order errors. By following this learning rule, synapses learn to correctly estimate the confidence of the associated prediction, which we use as a context-specific metric. Since $\bm{\pi_{\ell}}=\bm{A_{\ell}}\bm{r_{\ell+1}}$ approximates an inverse variance and enters as a metric in Eqs.~\ref{eq:E} and \ref{eq:udot}, it should remain positive. An important extension of Eq.~\ref{eq:Adot} is then to include a mechanism to ensure that components of $\bm{A_{\ell}}$ remain positive (see Methods). 

These two similar learning rules state that synaptic weights evolve to minimize errors remaining after inference. We verify in simulations that Eqs.~\ref{eq:Wdot} and \ref{eq:Adot} (with an additional mechanism to ensure positivity, see Methods) can indeed learn correct mean and confidence of different context-dependent data distributions as functions of higher-level representations (see SI5). Importantly, all the information needed for learning, namely the presynaptic rate and postsynaptic error, is readily available in the vicinity of the synapse. 

Having developed a way to learn how to estimate top-down confidence, we will now further examine how this is used in neuronal dynamics.

\subsection{Dynamic balancing of cortical streams based on confidence}
\begin{figure}[!ht]%
    \centering
    \includegraphics{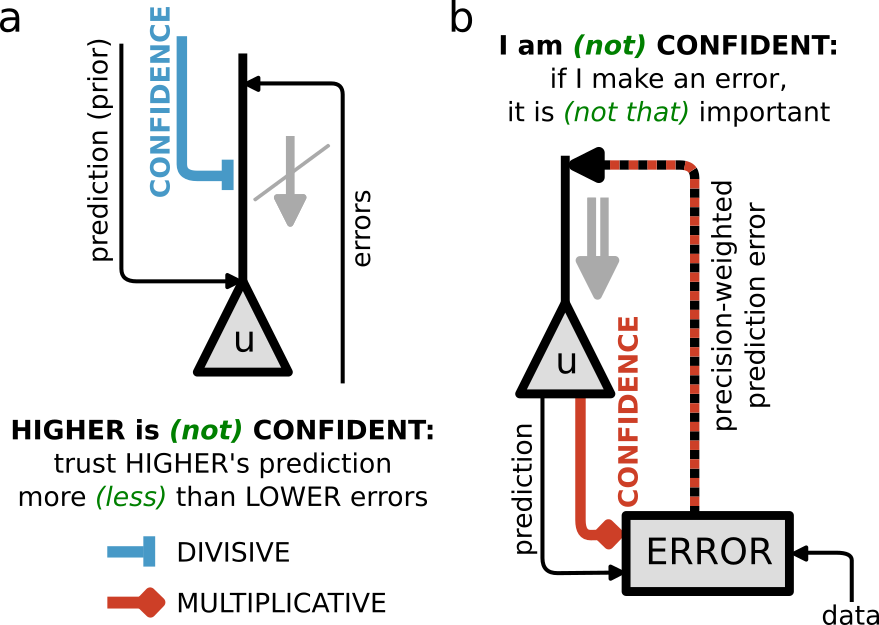}
    \caption{Adaptive balancing of cortical streams based on confidence.
    (a) Divisive weighting of errors by the confidence of top-down predictions about what the activity of a neuron should be (prior confidence, $\bm{\pi_{\ell}}^{-1}$).
    (b) Multiplicative weighting of errors by the confidence of predictions that a neuron makes about what the activity of other neurons should be (data confidence, $\bm{\pi_{\ell-1}}$).}\label{fig:balancing}
    \end{figure}

In our neuronal dynamics (Eqs. \ref{eq:udot} and \ref{eq:a}), the relative importance given to top-down predictions and bottom-up prediction errors is controlled by two mechanisms that both modulate the gain of prediction errors.
First, the confidence of top-down predictions of a neuron's activity (the prior) divisively impacts the importance of bottom-up errors in the inference dynamics of this neuron (see Fig.~\ref{fig:balancing}a).
For example, neurons encoding context might send more or less confident predictions to neurons encoding the presence of particular objects. Then, the relative importance of the prior prediction compared to bottom-up errors is greater in contexts sending more confident prior predictions (``In a forest, I know there are trees") than less confident ones (``In this city neighborhood, there might be trees, let's see...").

Second, the confidence of predictions a neuron makes about lower-level activities multiplicatively impacts the importance of errors entailed by these predictions (see Fig.~\ref{fig:balancing}b).
For example, neurons encoding object identity might send predictions to different sensory modalities with different confidence levels, reflecting different levels of reliability or noise in different lower-level streams. Prediction errors arising from more reliable streams should be weighted more strongly (``Across trees, structure (trunk, branches, leaves, etc.) is usually more consistent than color. To recognize a tree, I should then trust structure more than color").

This weighting is proportional to the more classical Bayes-optimal weighting of top-down prediction (akin to prior) and bottom-up errors (akin to data) by their respective reliabilities, and leads to a Bayes-optimal estimate of latent variables at equilibrium of neuronal dynamics.
Computationally, this mechanism proves valuable when integrating information from sources with different levels of reliability (or noise), for example, when integrating prior and data (see SI6) or during multimodal integration.

At the level of a cortical area, confidence controls the balance of bottom-up and top-down information on a neuron-by-neuron basis, providing fine-grained control over what is attended to.
It is worth highlighting that, with our formulation of confidence as a function of higher-level representations, we can encompass state-, context-, task- or feature-dependent confidence signals, depending on what the higher-level representations encode.
Moreover, as higher-level representations change, so do confidence signals, providing a mechanism to explain the trial-to-trial variability of confidence weighting observed in animals \cite{fetsch2009dynamic}.

At the level of a cortical area, confidence controls the balance of bottom-up and top-down information on a neuron-by-neuron basis, providing fine-grained control over what is attended to.
It is worth highlighting that, with our formulation of confidence as a function of higher-level representations, we can encompass state-, context-, task- or feature-dependent confidence signals, depending on what the higher-level representations encode.
Moreover, as higher-level representations change, so do confidence signals, providing a mechanism to explain the trial-to-trial variability of confidence weighting observed in animals \cite{fetsch2009dynamic}.

\subsection{Second-order error propagation}

\begin{figure*}[!ht]
\centering
\includegraphics{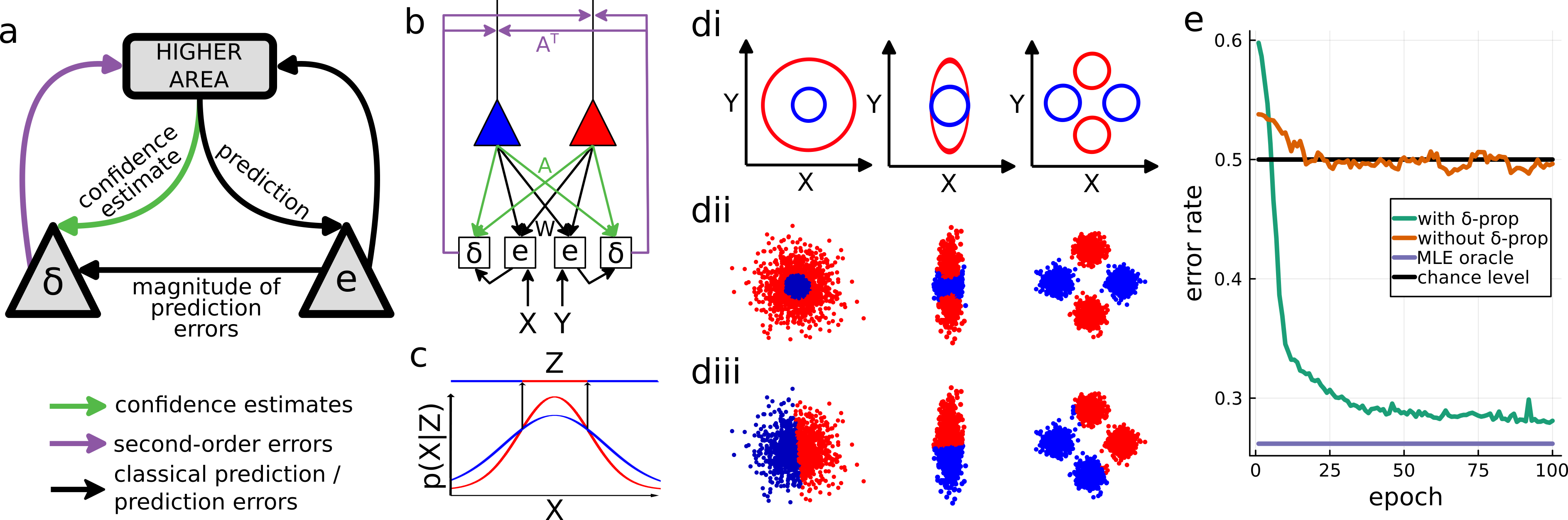}
\caption{Propagation of second-order errors for classification.
(a) Second-order errors compare confidence and performance (magnitude of prediction errors).
(b) A 2x2 network for binary classification. During learning, the $\mathrm{X}$ and $\mathrm{Y}$ data are sampled from one of the two class distributions, and the activity of neurons representing the class is clamped to the one-hot encoded correct class. Parameters ($\bm{W} ,\bm{A}$) are then learned following Eqs.~\ref{eq:Wdot} and \ref{eq:Adot}. During inference, the activity of neurons representing the class follows neuronal dynamics (without top-down influence), and we read the selected class as the one corresponding to the most active neuron. Prediction error (first-order) propagation is omitted in the depiction.
(c) Maximizing the likelihood of predictions leads to nonlinear classification in a single area. 
(di) Two different 2-dimensional binary classification tasks. The ellipse represents the true class distributions for the two classes.
(dii) Classification with second-order error propagation.
(diii) Classification without second-order error propagation.
(e) Classification accuracy on the task presented in d, second column.}\label{fig:deltaprop}
\end{figure*}

In the proposed neuronal dynamics (Eqs.~\ref{eq:udot} and \ref{eq:a}), second-order errors $\bm{\delta_{\ell}}$ are propagated through the cortical hierarchy alongside confidence-weighted prediction errors $\bm{\pi_{\ell}}\circ\bm{e_{\ell}}$ (see Fig.~\ref{fig:deltaprop}a). This entails a second-order cortical stream along which areas exchange confidence and second-order errors. Importantly, this means that the second-order errors change higher-level representations. 

To investigate the computational role of second-order error propagation and their influence on higher-level representations, we place a single area (a network without hidden layers, see Fig.~\ref{fig:deltaprop}b) in supervised learning settings on simple nonlinear binary classification tasks (see Fig~\ref{fig:deltaprop}di and Methods).
Parameters are learned following Eqs.~\ref{eq:Wdot} and \ref{eq:Adot}. As expected, the confidence signal after learning represents the class-specific inverse variance (see Fig.~S1).
With our dynamics (see Fig.~\ref{fig:deltaprop}dii), but not with classical predictive coding dynamics (see Fig.~\ref{fig:deltaprop}diii), a single area can solve these nonlinear classification tasks (see Fig.~\ref{fig:deltaprop}e). 

At a computational level, the qualitative difference in performance can be understood by looking at the energy we minimize.
With our model, we choose the latent representation which sends a predictive distribution with the highest likelihood with respect to current data (see Fig.~\ref{fig:deltaprop}c). In contrast, classical predictive coding chooses the latent representation that minimizes the Euclidean distance between the input and the entailed point prediction.
At an algorithmic level, the capacity of our network to solve these tasks comes from the influence of second-order errors on the higher-level representation.
To minimize second-order errors, the network must not only choose the class whose point (mean) prediction is closest to the data point (that is, first-order prediction error minimization). This is non-informative in the example in Fig.~\ref{fig:deltaprop} because both class distributions have the same mean. The network also has to choose the class that best predicts the remaining distance ($e_\ell^2$) between point prediction and data.

\subsection{Confidence estimation in cortical circuits}

We next describe how our dynamics could be realized in cortical circuits.
We postulate that latent variables $\bm{u_{\ell}}$ are encoded in the somatic activity of a population of intracortical pyramidal cells of layer 6 (L6p). As demanded by our theoretical framework, these neurons receive the majority of their input from intracortical long-range projections \cite{zolnik2020layer} and send top-down projections to lower cortical areas \cite{markov2014anatomy, rockland2019we}.
We propose that these projections carry not only predictions \cite{schwiedrzik2017high, schneider2018cortical, garner2022cortical}, but also confidence.
Following experimental evidence of error or mismatch encoding in pyramidal cells of cortical layer 2/3 \cite{zmarz2016mismatch, jordan2020opposing}, we propose that confidence-weighted prediction errors $\bm{\pi_{\ell}}\circ\bm{e_{\ell}}$ and second-order errors $\bm{\delta_{\ell}}$ are computed by two populations of pyramidal neurons situated in layer 3, respectively L3$e$ and L3$\delta$. As our theory demands, these neurons send feedforward projections to higher cortical areas \cite{markov2014anatomy, rockland2019we}.
Additionally, our theory suggests that both types of error are integrated into the total propagated errors $\bm{a_{\ell}}$ (as defined in Eq.~\ref{eq:a}). We propose that this integration takes place in distal apical dendrites of L6p situated at the height of layer 4/5a \cite{ledergerber2010properties}, in line with previous work postulating error encoding in segregated dendritic compartments \cite{sacramento2018dendritic}.

\begin{figure}[!t]%
\centering
\includegraphics{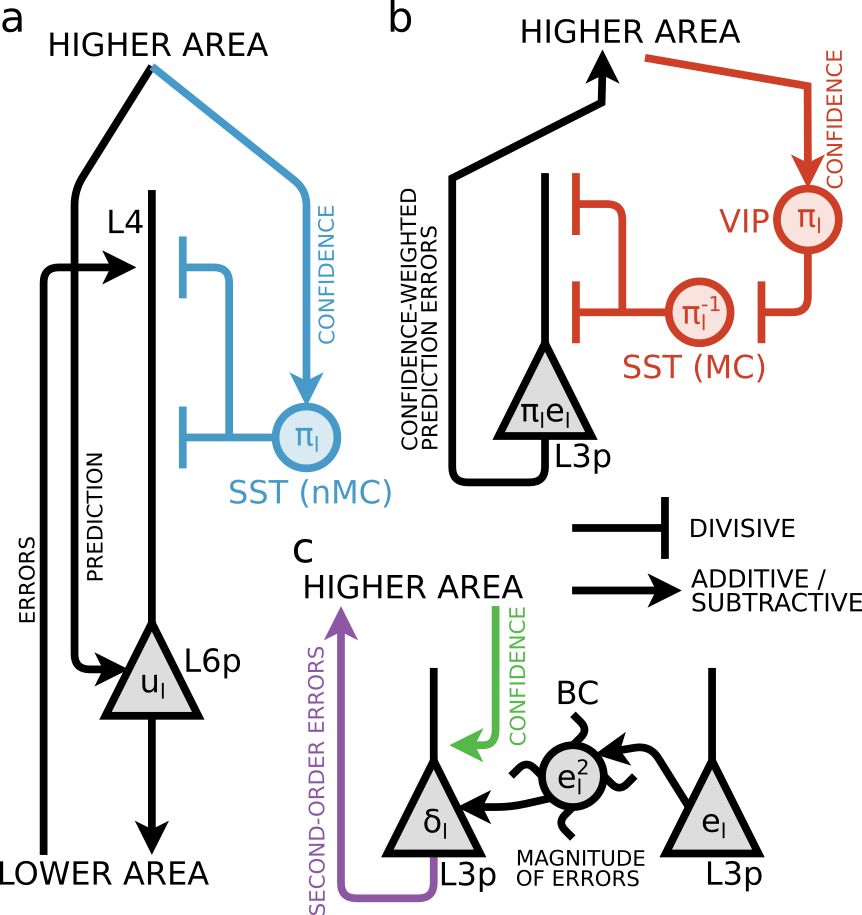}
\caption{Cortical circuit for neuronal dynamics of inference (as described in Eq.~\ref{eq:udot} and Eq.~\ref{eq:a}). 
(a) 
Representations ($\bm{u_{\ell}}$) are held in the somatic membrane potential of L6p.
Top-down synapses carrying predictions ($\bm{\mu_{\ell}}=\bm{W_{\ell}r_{\ell+1}}$) directly excite L6p at proximal dendrites.
Bottom-up confidence-weighted prediction errors ($\bm{W_{\ell-1}}^T(\bm{\pi_{\ell-1}} \circ \bm{e_{\ell-1}})$) and second-order errors ($\bm{A_{\ell-1}}^T\bm{\delta_{\ell-1}}$) are integrated into total error ($\bm{a_{\ell}}$) in the distal dendrites of L6p as described in Eq.~\ref{eq:a}.
This total error is then weighted by the prior uncertainty ($\bm{\pi_{\ell}}^{-1}$) through divisive dendritic inhibition realized by deep SST-expressing interneurons.
(b) 
Top-down predictions ($\bm{\mu_{\ell}}=\bm{W_{\ell}r_{\ell+1}}$) and local representations ($\bm{u_{\ell}}$) are compared in L3$e$.
Confidence weighting is then realized through gain modulation of L3$e$ by the disinhibitory VIP-expressing and SST-expressing interneurons circuit. 
(c) 
L3$\delta$ compares top-down confidence and local squared prediction errors encoded in basket cells (BC) into re-weighted second-order errors.}\label{fig:circuits}
\end{figure}

We now concern ourselves with the balancing of cortical streams through inhibition and disinhibition of errors entailed by our theory.
We propose that prediction errors are computed in L3$e$ by comparing local and top-down inputs from L6p.
Confidence weighting of bottom-up prediction errors might be realized through top-down gain modulation targeting L3$e$.
This could be achieved through a well-known disinhibitory circuit motif involving VIP-expressing interneurons receiving top-down input and preferentially inhibiting SST-expressing interneurons which in turn preferentially inhibit dendrites of L3$e$ \cite{pi2013cortical, lee2013disinhibitory, zhang2014long}.
This would entail VIPs encoding a confidence signal and superficial SSTs an expected uncertainty signal.
This hypothesis is corroborated by recent 2-photon imaging on rodents placed in an oddball paradigm, where activity ramps up in VIPs and decays in superficial SSTs as a stimulus is repeated \cite{bastos2023top}. 
Moreover, our theory suggests that total bottom-up errors should be modulated by the uncertainty of top-down predictions (the factor $\bm{\pi_{\ell}}^{-1}$ in Eq.~\ref{eq:udot}). In other words, at a circuit level, the confidence in prior, top-down information controls the integration of bottom-up errors by modulating the gain of somatic integration of apical activity. We propose that this is realized through modulation of L6p apical dendrites by deep (non-Martinotti) SST interneurons, which would then encode a confidence signal. 
The laminar specificity of SST activity \cite{munoz2017layer} and targets \cite{naka2019complementary} supports this hypothesis. 

Finally, we suggest circuit-level mechanisms underlying second-order error computation in L3$\delta$.
To compute second-order errors, confidence must be compared to the magnitude of current prediction errors.
We propose that the magnitude of prediction errors is computed in PV-expressing basket cells from local L3$e$ inputs.
At a circuit level, L3$e$ is thought to be separated into two populations encoding the positive and negative part of prediction errors, respectively \cite{jordan2020opposing}.
If this holds, then excitatory projections from both these populations to local basket cells, eventually followed by a nonlinear integration by basket cells \cite{cornford2019dendritic}, would be sufficient to perform the needed computation of local error magnitude \cite{hertag2023knowing}.
L3$\delta$ would then compute second-order errors by comparing top-down confidence and local (subtractive) inputs from basket cells. PV-expressing basket cells have indeed been shown to preferentially inhibit specific pyramidal cell types \cite{lee2014pyramidal, schneider2023cell}. Finally, recent evidence suggests that L3$e$ expresses \textit{Adamts2} and \textit{Rrad} \cite{o2023molecularly}, while no functional role has yet been proposed for the third class of superficial pyramidal cells expressing \textit{Agmat}, which we propose could be L3$\delta$.
The here presented propositions can serve as a starting point for experimental investigation of cortical second-order errors.

\section{Discussion}

In this work, we derived predictive coding dynamics with adaptive, context-dependent and learned confidence information. 
Specifically, we considered diagonal estimates of the inverse covariance matrix (with diagonal $\bm{\pi_{\ell}}$). In that case, each input dimension is scaled by the corresponding standard deviation when computing distances.
However, the brain's utilization of (inverse) variance estimates is likely to encompass various forms beyond the diagonal estimates explored in our study. 
Scalar estimates would define the importance granted to all errors in an area. In that case, confidence weighting of errors might be realized through nonspecific release of neuromodulators \cite{angela2005uncertainty, lawson2021computational}, scaling all feature dimensions equally.
On the other end of the spectrum, we might consider full inverse covariance matrices.
We would then consider not only stretch but also skew in our metric.
Doing so might lead to a theoretically grounded account of lateral connections between prediction error nodes \cite{friston2005theory}, with links to the notion of statistical whitening (the matrix square root of the full inverse covariance matrix is the ZCA whitening matrix). In general, we emphasize that considering the variance of predictive distributions as the backbone for normative theories of cortical modulation seems to us a promising endeavor. 

Moreover, we treated predictions (mean estimates) as arising in a top-down manner, supposing that the cortex performs inference and learning on a purely generative model of its inputs. Considering bottom-up instead of top-down predictions is mathematically straightforward, could potentially align better with models considering top-down cortical pathways as modulating activity in a feedforward feature detection stream \cite{gilbert2013top, klink2017distinct}, and could facilitate more direct testing on discriminative machine learning benchmarks. At the level of cortical circuitry, one might consider forward-projecting layer 5a and backward-projecting layer 2/3a pyramidal cells as sending bottom-up predictions and the entailed top-down prediction errors, respectively. This would form a fundamentally discriminative cortical pathway, a sort of dual of the generative pathway considered in this work.

It is important to acknowledge that the dynamics that we presented in this work share some classical limitations of predictive coding dynamics concerning biological plausibility e.g., weight transport \cite{max2023learning}, long inference \cite{haider2021latent}, encoding of signed errors \cite{sacramento2018dendritic, jordan2020opposing}, one-to-one connections, weak criteria of locality for learning and the assumption of a strict hierarchy of latent variables \cite{suzuki2023deep}. 

In our model, confidence multiplicatively modulates errors and is computed at each level of the hierarchy as a function of current representations.
This dynamic gain modulation is reminiscent of the attentional mechanism in transformer networks \cite{vaswani2017attention}. Our formulation offers a first step towards a bridge between models of attention in terms of neural gain modulation based on confidence \cite{feldman2010attention, kok2012attention, jiang2013attention} and attentional mechanisms in machine learning.
Anecdotally, VIP-expressing interneurons, encoding confidence in our model, were described by experimentalists as ``generating a spotlight of attention" \cite{karnani2016opening}.
Furthermore, the computational interest of dynamic top-down gain modulation might also be sought through the lens of efficient and parsimonious coding \cite{mlynarski2022efficient}. 
This perspective may already be implicitly embedded within our framework, given the connection between maximum likelihood and the infomax principle \cite{cardoso1997infomax}.

Confidence weighting of prediction errors occurs as a central element in leading models of psychopathologies under the predictive processing framework \cite{van2014precise, sterzer2018predictive, corlett2019hallucinations, friston2022computational}. These models are often based on the idea of a pathological (over- or under-) weighting of either prior or data in a process of Bayesian integration. In our model, these two hypotheses involve distinct neural mechanisms, that is, modulation, respectively, of L6p apical dendrites and L3p. This distinction might prove critical to extending these models to the whole cortical hierarchy, where activity at one level both represents data for the level above and generates priors for the level below.
Moreover, our proposed computational roles for interneuron circuitry might help link accounts of neuropsychiatric disorders in terms of confidence weighting of errors to accounts in terms of cortical excitation-inhibition balance \cite{sohal2019excitation} and interneuron dysfunction \cite{marin2012interneuron}.


It has previously been suggested that prediction error responses of layer 2/3 cells should be modulated by the expected uncertainty of the predicted feature\cite{wilmes2023uncertainty}. Our derivation suggests that the same prediction errors should in addition be weighted by the expected uncertainty of the feature generating the prediction. Accordingly, we propose the terminology of doubly uncertainty-modulated prediction errors.

Finally, the suggested implementation in the circuitry of cortical pyramidal cells and interneurons definitely requires further refinement through experimental work.
Nevertheless, we provide a rigorous theoretical framework to interpret existing experimental results and formulate ideas for experimental testing. Beyond providing a specific set of predictions, we aim to convey a novel normative perspective which indicates that searching for signatures of confidence estimation and second-order errors in cortical circuits might be an interesting venue, especially in interneuron activity.

\section{Materials and Methods}
\small

\subsection{Probabilistic model}~
Here we elaborate on the form of the probabilistic model. We introduce a notion of strict hierarchy between levels of latent representations by supposing that the joint can be decomposed as 
\begin{equation}\label{eq:assumption1}
p(\bm{u_0},\bm{u_1}, \dots, \bm{u_n}) \propto p(\bm{u_0}|\bm{u_1})p(\bm{u_1}|\bm{u_2})\dots p(\bm{u_{n-1}}|\bm{u_n})\;,
\end{equation}
which can be justified by assuming a Markov property $\forall l \in [0,n), ~ p(\bm{u_{\ell}}|\bm{u_{\ell+1}}, \dots, \bm{u_n}) = p(\bm{u_{\ell}}|\bm{u_{\ell+1}})$ and a uniform top level prior $\bm{u_n} \sim \mathcal{U}$.
Since the distribution of $\bm{u_{\ell}}$ is conditioned on $\bm{u_{\ell+1}}$, we call this a generative hierarchy. 

We further assume that predictions  $\bm{u_{\ell}}|\bm{u_{\ell+1}}$ follow multivariate Gaussian distributions
\begin{equation}\label{eq:assumption2}
\bm{u_{\ell}}|\bm{u_{\ell+1}} \sim \mathcal{N}\left(\bm{W_{\ell}}\bm{r_{\ell+1}}, \text{diag}(\bm{A_{\ell}}\bm{r_{\ell+1}})^{-1} \right)
\end{equation} 
with mean at point predictions $\bm{W_{\ell}}\bm{r_{\ell+1}}$ and diagonal covariance matrix with diagonal $(\bm{A_{\ell}r_{\ell+1}})^{-1}$.

Under the two assumptions described in Eqs.~\ref{eq:assumption1} and \ref{eq:assumption2}, we have 
\begin{equation}
-\log p(\bm{u_0},\bm{u_1}, \dots, \bm{u_n}) + \mathrm{const} = E\;.
\end{equation} 

\subsection{Confidence as metrics in neuronal dynamics}~
In this work, we chose confidence as a metric for neuronal dynamics Eq.~\ref{eq:udot} (see \cite{surace2020choice} for an introduction to the use of metrics in gradient-based dynamics in neuroscience).
Note that if we make the approximation of considering that predictions are fixed during inference (a ``fixed-prediction assumption" \cite{millidge2022predictive}), the confidence is the second derivative of the energy.
Second derivatives provide additional information on the curvature of the energy landscape and are known to have desirable properties as metrics (second-order optimization). A striking limitation, however, lies in assuming fixed predictions during inference, the confidence is only a crude approximation of the actual second derivative without fixed predictions (see SI2 for the actual second derivative).
An intuition of the effect of this change of metric on neuronal dynamics (Eq.~\ref{eq:udot}) is as normalizing the balance of importance between local and lower prediction errors such that the importance of local errors is 1. 

\subsection{Positivity of confidence}
A sufficient condition for neuronal dynamics Eq.~\ref{eq:udot} to follow a descent direction on $E$ is that all terms of $\bm{\pi_{\ell}}=\bm{A_{\ell}}\bm{r_{\ell+1}}$ are positive. Let us assume that rates $\bm{r_{\ell+1}}$ are positive (the neuronal transfer function $\phi$ outputs positive values). Then a sufficient condition is that all components of $\bm{A_{\ell}}$ also are positive. There are multiple possible extensions of Eq.~\ref{eq:Adot} to enforce this. One is to initialize all components of $\bm{A_{\ell}}$ to positive values and to modulate the learning rate by the current weights  
\begin{equation}
    \dot{\bm{A_{\ell}}} \propto \bm{A_{\ell}} \circ \bm{\delta_{\ell}}\bm{r_{\ell+1}}^T\;,  
\end{equation} essentially preventing weights from crossing $0$. This is necessary to stabilize learning when scaling up to more complex settings (see SI5). This is also in accordance with the general physiological fact that the sign of synaptic influence cannot change.

\subsection{Simulation details}~
For simulations presented in Fig.~\ref{fig:deltaprop}, we built the datasets by sampling $N=1000$ points $(x_1, y_1), \dots, (x_N, y_N)$ from each of the data distributions represented in Fig.~\ref{fig:deltaprop}di (by their 99.7\% confidence ellipses), and attaching the corresponding class label (either red or blue). 
We then build a 2x2 network where the top level activity is a one-hot representation of the class label and the bottom level activity is the coordinate in space $(x,y)$.
We train this network in supervised learning settings on the dataset by clamping both the top and bottom areas to the corresponding elements of the dataset and perform one step of parameters learning as described in Eqs.~\ref{eq:Wdot} and \ref{eq:Adot}.

We then test the capacity of our network to classify data by only clamping the bottom level to the data and letting the top-level activity follow Eq.~\ref{eq:udot}.
We select as the output class index the index of the maximum top-level activity and plot the corresponding classification in Fig.~\ref{fig:deltaprop}dii. 
For comparison, we also plot in Fig.~\ref{fig:deltaprop}diii the classification results obtained with the same 2x2 architecture but using classical predictive coding dynamics and following the same training and testing procedures.
In Fig.~\ref{fig:deltaprop}e we plot the associated performance, with the addition of the maximum likelihood estimate with perfect knowledge of the means and variances.

Simulations and pseudocodes for confidence learning and Bayes-optimal integration in dynamic environments are reported in SI5 and SI6.

\section{Acknowledgments}
This work has received funding from the European Union 7th Framework Programme under grant agreement 604102 (HBP), the Horizon 2020 Framework Programme under grant agreements 720270, 785907 and 945539 (HBP) and the Manfred Stärk Foundation. 
\section{Code availability}
Simulation code for this paper can be accessed at \url{github.com/arnogranier/precision-estimation}.
\section{Author contributions}
A.G. conceptualized the outlines of the paper, compiled and organized literature materials, performed the simulations and wrote the original draft. A.G. and W.S. participated in the derivation and presentation of the mathematical formalism. K.W. and A.G. had the original idea and K.W. supervised the work together with W.S. All authors participated in conceptualization, interpretation of results, reviewing and editing of the manuscript and approved the final manuscript.
\section{Competing interests}
The authors declare no competing interests.
\printbibliography

\newpage
\beginsupplement

\section*{Supplementary Information}
\setcounter{section}{0}

\section{Energy}

The density of a multivariate Gaussian with diagonal covariance $\bm{\Sigma} = \text{diag}(\bm{\sigma}^2)$ with $\bm{\sigma}^2>0$ is 
\begin{align}
f(\bm{u};\bm{\mu}, \bm{\sigma}^2) &= (2\pi)^{-k/2}\text{det}(\bm{\Sigma})^{-1/2}\exp\left(-\frac12(\bm{u}-\bm{\mu})^T\bm{\Sigma}^{-1}(\bm{u}-\bm{\mu})\right) \label{eq:multi}
\\&= (2\pi)^{-k/2}\left(\prod_i\bm{\pi}_i\right)^{1/2}\exp\left(-\frac12{\|\bm{e}\|}_{\bm{\pi}}^2\right)\;,\label{eq:multisimple}
\end{align}
noting $\bm{\pi} = \bm{1}/\bm{\sigma}^2$ where the division is taken elementwise and ${\|\bm{e}\|}_{\bm{\pi}}^2 = \|\bm{u}-\bm{\mu}\|^2_{\bm{\Sigma}^{-1}} = (\bm{u}-\bm{\mu})^T\bm{\Sigma}^{-1}(\bm{u}-\bm{\mu})$.
For the determinant, remark that the determinant of diagonal matrix is the product of its diagonal elements.\\
We now derive Eq.~8
\begin{align}
-\log p(\bm{u_0},\bm{u_1},\dots,\bm{u_n}) &= -\log\left(K\prod_{l=0}^{n-1} p(\bm{u_{\ell}}|\bm{u_{\ell+1}})\right) \label{eq:step1}\\
&= - \sum_{l=0}^{n-1} \log p(\bm{u_{\ell}}|\bm{u_{\ell+1}}) + K\\
&= - \sum_{l=0}^{n-1} \log\left((2\pi)^{-k_{\ell}/2}\left(\prod_i(\bm{\pi_{\ell}})_i\right)^{1/2}\exp\left(-\frac12{\|\bm{e_{\ell}}\|}_{\bm{\pi_{\ell}}}^2\right)\right) + K\label{eq:step3}\\
&= - \sum_{l=0}^{n-1} \log\left((2\pi)^{-k_{\ell}/2}\right) - \frac12\sum_{l=0}^{n-1} \log\left(\prod_i(\bm{\pi_{\ell}})_i\right) + \frac12\sum_{l=0}^{n-1} {\|\bm{e_{\ell}}\|}_{\bm{\pi_{\ell}}}^2 + K\\
&= \frac12 \sum_{l=0}^{n-1} {\|\bm{e_{\ell}}\|}_{\bm{\pi_{\ell}}}^2 - \frac12 \sum_{l=0}^{n-1} \log|\bm{\pi_{\ell}}| + K\;,
\end{align}
where to get Eq.~\ref{eq:step1} we used Eq.~6 and to get Eq.~\ref{eq:step3} we used Eqs.~7 and \ref{eq:multisimple}.

\section{Partial derivatives of the energy}
We now give a high-level view of the derivation of partial derivatives of the energy $E$ used in neuronal and synaptic dynamics Eqs.~2, 4 and 5. We omit calculation details for the sake of brevity. As a reminder, we set $\bm{e_{\ell}} = \bm{u_{\ell}} - \bm{W_{\ell}}\phi(\bm{u_{\ell+1}})$, $\bm{\pi_{\ell}} = \bm{A_{\ell}}\phi(\bm{u_{\ell+1}})$, $\bm{\delta_{\ell}} = (\bm{\pi_{\ell}}^{-1}-\bm{e_{\ell}}^2)/2$ and $\circ$ is the componentwise (Hadamard) product. \\
For this, we will make use of the following matrix calculus formulas:
\begin{equation}
    \forall \bm{M} \text{ symmetric },\frac{\partial \bm{x}^T\bm{M}\bm{x}}{\partial \bm{x}} = 2\bm{M}\bm{x}\;, \tag{i} \label{eq:i}
\end{equation}
\begin{equation}
    \frac{\partial \bm{g(x)}}{\partial \bm{x}} = \frac{\partial g(f(\bm{x}))}{\partial f(\bm{x})}\frac{\partial f(\bm{x})}{\partial \bm{x}} ~~\text{(chain rule)}\;,\tag{ii} \label{eq:ii}
\end{equation}
\begin{equation}
    \frac{\partial \bm{1}^T\log(\bm{M}\bm{x})}{\partial \bm{x}} = \bm{1}^T(\text{diag}(\bm{1}/(\bm{Mx}))\bm{M}) = \bm{M}^T(\bm{1}/(\bm{Mx})) \text{ (with the division being componentwise) }\;,\tag{iii} \label{eq:iii}
\end{equation}
\begin{equation}
    \frac{\partial f(\bm{x})^Tg(\bm{x})}{\partial \bm{x}} = \frac{\partial f(\bm{x})}{\partial \bm{x}}g(\bm{x})+\frac{\partial g(\bm{x})}{\partial \bm{x}}f(\bm{x})\;,\tag{iv} \label{eq:iv}
\end{equation}
\begin{equation}
    \frac{\partial f(\bm{x}) \circ g(\bm{x})}{\partial \bm{x}} = \frac{\partial f(\bm{x})}{\partial \bm{x}}\text{diag}(g(\bm{x}))+\frac{\partial g(\bm{x})}{\partial \bm{x}}\text{diag}(f(\bm{x}))\;.\tag{v} \label{eq:v}
\end{equation}

\subsubsection*{Latent variables}
The derivative with respect to $\bm{u_{\ell}}$ can be decomposed in three terms
\begin{equation}
    2\frac{\partial E}{\partial\bm{u_{\ell}}} = \frac{\partial \|\bm{e_{\ell}}\|_{\bm{\pi_{\ell}}}^2}{\partial \bm{u_{\ell}}} + \frac{\partial \|\bm{e_{\ell-1}}\|_{\bm{\pi_{\ell-1}}}^2}{\partial \bm{u_{\ell}}} - \frac{\partial \log|\bm{\pi_{\ell-1}}|}{\partial \bm{u_{\ell}}}\;.
\end{equation}
We compute those three terms independently.\\
For the first term, the derivation is straightforward and follow directly from (\ref{eq:i}) and (\ref{eq:ii}) \begin{equation}
    \frac{\partial \|\bm{e_{\ell}}\|_{\bm{\pi_{\ell}}}^2}{\partial \bm{u_{\ell}}} = 2\bm{\pi_{\ell}} \circ \bm{e_{\ell}}\;.
\end{equation}
For the second term we first remark that it can be written as $\frac{\partial \bm{e_{\ell-1}}^T(\bm{\pi_{\ell-1}} \circ \bm{e_{\ell-1}})}{\partial \bm{u_{\ell}}}$, apply (\ref{eq:iv}) and then develop $\frac{\partial \bm{\pi_{\ell-1}} \circ \bm{e_{\ell-1}}}{\partial \bm{u_{\ell}}}$ following (\ref{eq:v})
\begin{equation}
    \frac{\partial \|\bm{e_{\ell-1}}\|_{\bm{\pi_{\ell-1}}}^2}{\partial \bm{u_{\ell}}} = -2\phi'(\bm{u_{\ell}})\circ\left(\bm{W_{\ell-1}}^T(\bm{\pi_{\ell-1}} \circ \bm{e_{\ell-1}}) - \frac12\bm{A_{\ell-1}}^T\bm{e_{\ell-1}}^2\right)\;.
\end{equation}
For the third term remark that $\log |\bm{x}|=\bm{1}^T\log \bm{x}$, then a straightforward application of (\ref{eq:iii}) is sufficient
\begin{equation}
\frac{\partial \log|\bm{\pi_{\ell-1}}|}{\partial \bm{u_{\ell}}} = \phi'(\bm{u_{\ell}}) \circ \bm{A_{\ell-1}}^T\bm{\pi_{\ell-1}}^{-1}\;.
\end{equation}Finally putting it all together we have
\begin{equation}
    \frac{\partial E}{\partial\bm{u_{\ell}}} = \bm{\pi_{\ell}} \circ \bm{e_{\ell}} - \phi'(\bm{u_{\ell}}) \circ \left(\bm{W_{\ell-1}}^T(\bm{\pi_{\ell-1}} \circ \bm{e_{\ell-1}}) +\bm{A_{\ell-1}}^T\bm{\delta_{\ell-1}}\right)\;.
\end{equation}

\subsubsection*{Prediction weights}
The derivative with respect to $\bm{W_{\ell}}$ is simply
\begin{equation}
    2\frac{\partial E}{\partial\bm{W_{\ell}}} = \frac{\partial \|\bm{e_{\ell}}\|_{\bm{\pi_{\ell}}}^2}{\partial \bm{W_{\ell}}}\;.
\end{equation}
The derivation is straightforward and follow directly from (\ref{eq:i}) and (\ref{eq:ii})
\begin{equation}
    \frac{\partial \|\bm{e_{\ell}}\|_{\bm{\pi_{\ell}}}^2}{\partial \bm{W_{\ell}}} = \frac{\partial \bm{e_{\ell}}^T\text{diag}(\bm{\pi_{\ell}})\bm{e_{\ell}}}{\partial \bm{e_{\ell}}}\frac{\partial \bm{e_{\ell}}}{\partial \bm{W_{\ell}}} = -2(\bm{\pi_{\ell}} \circ \bm{e_{\ell}}) \phi(\bm{u_{\ell+1}})^T
\end{equation}
and
\begin{equation}
    \frac{\partial E}{\partial\bm{W_{\ell}}} = -(\bm{\pi_{\ell}} \circ \bm{e_{\ell}}) \phi(\bm{u_{\ell+1}})^T\;.
\end{equation}
\subsubsection*{Confidence estimation weights}
The derivative with respect to $\bm{W_{\ell}}$ can be decomposed in two terms
\begin{equation}
    2\frac{\partial E}{\partial\bm{A_{\ell}}} = \frac{\partial \|\bm{e_{\ell}}\|_{\bm{\pi_{\ell}}}^2}{\partial \bm{A_{\ell}}} - \frac{\partial\log|\bm{\pi_{l}}|}{\partial \bm{A_{\ell}}}\;.
\end{equation} 
We compute those two terms independently. For these we find it easier to compute derivatives element by element. \\
For the first term remark that $\|\bm{e_{\ell}}\|_{\bm{\pi_{\ell}}}^2 = \sum_i \left({\bm{e_{\ell}}}^2\right)_i \sum_j \left(\bm{A_{\ell}}\right)_{i,j} \left(\phi(\bm{u_{\ell+1}})\right)_j$ and then it is simple to see that 
\begin{equation}
    \frac{\partial \|\bm{e_{\ell}}\|_{\bm{\pi_{\ell}}}^2}{\partial (\bm{A_{\ell}})_{i,j}} = \left({\bm{e_{\ell}}}^2\right)_i\left(\phi(\bm{u_{\ell+1}})\right)_j\;.
\end{equation} 
For the second term remark that $\log|\bm{\pi_{l}}| = \sum_i \log (\bm{\pi_{\ell}})_i = \sum_i\log\left(\sum_j  (\bm{A_{\ell}})_{i,j}(\phi(\bm{u_{\ell+1}}))_j\right)$, and then
\begin{equation}
    \frac{\partial\log|\bm{\pi_{l}}|}{\partial (\bm{A_{\ell}})_{i,j}} = \frac{\left(\phi(\bm{u_{\ell+1}})\right)_j}{\sum_j (\bm{A_{\ell}})_{i,j}\left(\phi(\bm{u_{\ell+1}})\right)_j} = \left(\bm{\pi_{\ell}}^{-1}\right)_i\left(\phi(\bm{u_{\ell+1}})\right)_j\;.
\end{equation}
Putting it together and writing it in matrix form
\begin{equation}
    \frac{\partial E}{\partial\bm{A_{\ell}}} = -\bm{\delta_{\ell}}\phi(\bm{u_{\ell+1}})^T\;.
\end{equation}
\subsubsection*{Latent variables - second derivative}
The second derivative of the energy with respect to latent representations is
\begin{align*}\label{eq:de2}
    \frac{\partial^2 E}{\partial\bm{u_{\ell}}^2} = &\text{diag}(\bm{\pi_{\ell}}) - \text{diag}(\bm{r_{\ell}}''\circ(\bm{W_{\ell-1}}^T(\bm{\pi_{\ell-1}} \circ \bm{e_{\ell-1}}) +  \bm{A_{\ell-1}}^T\bm{\delta_{\ell-1}})) \\ &+ \text{diag}(\bm{r_{\ell}}')(\bm{W_{\ell-1}}^T\text{diag}(\bm{e_{\ell-1}})\bm{A_{\ell-1}} - \bm{W_{\ell-1}}^T\text{diag}(\bm{\pi_{\ell-1}})\bm{W_{\ell-1}} \\ &~~~~~~~~~~~~~~~-0.5\bm{A_{\ell-1}}^T\text{diag}(\bm{\pi_{\ell-1}}^{-2})\bm{A_{\ell-1}}+\bm{A_{\ell-1}}^T\text{diag}(\bm{e_{\ell-1}})\bm{W_{\ell-1}}^T ) \numberthis
\end{align*}

\section{Metrics for parameter learning}
In this work, we took confidence as a metric when deriving neuronal dynamics (Eq.~2) but used the default Euclidean metric when deriving synaptic learning rules (Eqs.~4 and 5). 
The same approach of taking as a metric an approximate second-order derivative in gradient descent could be used not only for inference but also (and in fact more classically) for parameter learning. In that case, second derivatives are also expressed with confidence/variance:
\begin{equation}
    - \frac{\partial^2 \log f(\bm{u} ; \bm{m}, \bm{p}^{-1})}{\partial {\bm{m}}^2} = \text{diag}(\bm{p})\;,
\end{equation}
\begin{equation}
    - \frac{\partial^2 \log f(\bm{u} ; \bm{m}, \bm{p}^{-1})}{\partial {\bm{p}}^2} = \text{diag}(\bm{p}^{-2})\;.
\end{equation}
with $f$ being the density of a multivariate Gaussian with mean $\bm{m}$ and variance $\bm{p}^{-1}$ and $\bm{m}, \bm{p}$ not functions of $\bm{u}$.\\
Note that this is not always expressed correctly in the literature (e.g., \cite{millidge2021predictive, ofner2021predictive} Eq.~64, where the Fisher information matrix $\cal G$ instead of its inverse appears in the definition of natural gradient, c.f.~our Eq.~8), leading to confusion on the link between confidence weighting and natural gradient descent.

\section{Intuition at equilibrium in the linear case} \label{SI:bayes}

At equilibrium of Eq.~2, noting $\bm{\Pi_k} = \text{diag}(\bm{\pi_k})$, ignoring second-order errors ($\bm{\delta_{\ell-1}} = 0$) and working in the linear case $\phi(\bm{x})=\bm{x}$ we have the value at equilibrium
\begin{equation}
    \bm{u_{\ell}^*} = (\bm{\Pi_{\ell}}+\bm{W_{\ell-1}}^T\bm{\Pi_{\ell-1}}\bm{W_{\ell-1}})^{-1}(\bm{\Pi_{\ell}}\bm{W_{\ell}}\bm{u_{\ell+1}} + \bm{W_{\ell-1}}^T\bm{\Pi_{\ell-1}}\bm{u_{\ell-1}})\;,
\end{equation}
where the first term can be interpreted as a normalization factor and the second term as a weighted sum of higher and lower representations ``translated in the language'' of the local level $l$ through prediction weight matrices.
Remark that, if the confidence $\bm{\Pi_{\ell-1}}$ of the prediction that level $l$ makes about level $l-1$ is negligible compared to the confidence $\bm{\Pi_{l}}$ of the prediction that level $l+1$ makes about level $l$, which we will note $\bm{\Pi_{\ell-1}}/\bm{\Pi_{l}} \rightarrow 0$, then the activity of level $l$ goes to the prediction made by level $l+1$ (the prior)
\begin{equation}\label{eq:equi1}
    \bm{\Pi_{\ell-1}}/\bm{\Pi_{l}} \rightarrow 0 \;\Longrightarrow\; \bm{u_{\ell}^*} \rightarrow \bm{W_{\ell}}\bm{u_{\ell+1}}\;.
\end{equation}
Inversely, when the prediction that level $l+1$ makes about what the activity in level $l$ should be (the prior) is deemed unreliable compared to the prediction that level $l$ makes about what the activity of level $l-1$ should be, then the activity of level $l$ goes to a value such that its prediction is the activity in level $l-1$
\begin{equation}\label{eq:equi2}
    \bm{\Pi_{l}}/\bm{\Pi_{\ell-1}} \rightarrow 0 \;\Longrightarrow\; \bm{W_{\ell-1}}\bm{u_{\ell}^*} \rightarrow \bm{u_{\ell-1}}\;.
\end{equation}

\section{Simulation details: Confidence learning}

\begin{figure}[!ht]
\centering
\includegraphics{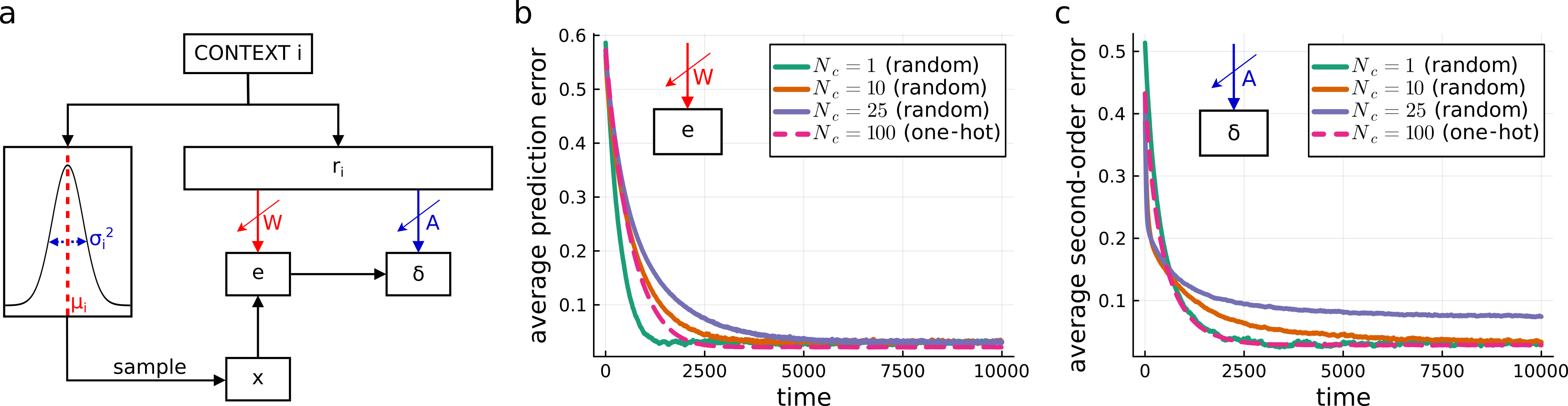}
\caption{Error-correcting synaptic learning. 
        (a) In these simulations, we consider a higher area with $N_{\ell+1}$ neurons and a lower area with $N_{\ell}$ neurons.
            Specifically, here we take $N_{\ell+1}=N_{\ell}=100$.
            The activity vector in the higher area can take $N_c$ different values [$\bm{r_n},~n\!=\!1,\dots,N_c$], to each of which is associated a different mean [$\bm{\mu_n}$] and a different variance [$\bm{\sigma_n}^2$].
            The activity in the lower area is then sampled from the Gaussian distribution with this mean and variance.
            Predictions [$\bm{Wr_i}$] and confidence estimates [$\bm{Ar_i}$] are formed from the higher-level representation and prediction errors [$\bm{e}=\bm{x}-\bm{Wr_i}$] and second-order errors [$\bm{\delta}=\bm{1}/\bm{Ar_i}-\bm{e}^2$] are computed and used to learn parameters [$\bm{W}$ and $\bm{A}$].
            For simulations marked \texttt{(random)}, higher-level representations are random binary vectors with an average of 50\% of ones.
            For simulations marked \texttt{(one-hot)}, higher-level representations are one-hot encoded.
        (b) Here we show that with the learning rule Eq.~4 the network correctly learns to estimate the means [$\bm{\mu_n},~n\!=\!1,\dots,N_c$] from higher-level activity [$\bm{r_n},~n\!=\!1,\dots,N_c$].
                In these simulations we suppose that the confidence estimate is $1$.
        (c) Here we show that with the learning rule Eq.~5 the network correctly learns to estimate the confidences [$\bm{1}/\bm{\sigma_n}^2$] from higher-level activity [$\bm{r_n}$].}
        \label{fig:learning}
\end{figure}

For simulations presented in Fig.~\ref{fig:learning}c, we follow the simulation setup presented in Fig.~\ref{fig:learning}a and described in more details below and in Supplementary Algorithm \ref{alg:Adot}.

We consider a higher area with $N_{\ell+1}$ neurons and a lower area with $N_{\ell}$ neurons. We consider $N_c$ different classes of inputs, each with its own distribution $\mathcal{N}(\bm{\mu_i}, \bm{\sigma_i}^2), i \in [1, N_c]$, where $\bm{\mu_i}$ and $\bm{\sigma_i}^2$ are vectors of size $N_{\ell}$. We initialize all $\bm{\mu_i}$ following a $\mathcal{U}(-1,1)$ and all $\bm{\sigma}^2_i$ following a $\mathcal{U}(1/4,1)$. Then we choose the representational mode of the higher area, either random binary vectors or one-hot encoded and initialize higher-level representations $\bm{r_i}, i \in [1,N_c]$ as random binary vectors of size $N_{\ell+1}$ with on average $p$ ones or one-hot encoded $i$ in $N_{\ell+1}$, respectively. The confidence estimation matrix $\bm{A}$ is then initialized as a matrix filled with $\alpha$, with $\alpha=1/pN_{\ell+1}$ for the random binary vector case and $\alpha=1$ for the one-hot encoded case. We then repeat the following procedure for multiple epochs: 
\begin{enumerate}
    \item For each class, sample a data point $\bm{x_i}$ from $\mathcal{N}(\bm{\mu_i}, \bm{\sigma_i}^2)$.
    \item Set the higher-level representation to $\bm{r_i}$.
    \item Compute the confidence estimate $\bm{\pi_i} = \bm{Ar_i}$.
    \item Compute the second-order error $\bm{\delta_i} = (1/\bm{\pi_i}-(\bm{x_i}-\bm{\mu_i})^2)/2$.
    \item Update $\bm{A}$ following Eq.~9.
\end{enumerate}
In Fig.~\ref{fig:learning}c, we plot the evolution of $(\sqrt{N_{\ell}}N_c)^{-1}\sum_i \|\bm{\sigma_i}^2 - 1/\bm{Ar_i}\|$ through epochs.
For Fig.~\ref{fig:learning}c, parameters are $T=10000, N_{\ell+1}=N_{\ell}=100, \eta=0.001$ with $N_c$ varying depending on the simulation.
A similar procedure is used for Fig.~\ref{fig:learning}b, but following Eq.~4.\\

\begin{breakablealgorithm}
\caption{Confidence learning}\label{alg:Adot}
\begin{algorithmic}
\Require $T$, $N_{\ell+1}$ $N_{\ell}$, $N_c$, $\eta$, overlap, $p$
\State $\bm{\sigma}^2 = [1.5~\text{rand}(N_{\ell}) + 0.5 \text{ for \_ in 1:}N_c]$ \Comment{$\bm{\sigma}^2$ initialization, random uniform between $1/2$ and $2$}
\State $\bm{\mu} = [2~\text{rand}(N_{\ell}) - 1 \text{ for \_ in 1:}N_c]$ \Comment{$\bm{\mu}$ initialization, random uniform between $-1$ and $1$}
\If{overlap}
\State $\bm{r} = [\text{rand}(N_{\ell+1}) < p \text{ for \_ in 1:}N_c]$ \Comment{$\bm{r}$ initialization, random binary vector with $p$\% ones on average}
\State $\bm{A} = \text{ones}(N_{\ell},N_{\ell+1})/(pN_{\ell+1})$ \Comment{$\bm{A}$ initialization (such that the mean starting $\bm{\pi}$ is one)}
\Else
\State $\bm{r} = [[\text{(j==i) ? 1 : 0 for j in 1:}N_{\ell+1}] \text{ for i in 1:}N_c]$ \Comment{$\bm{r}$ initialization, onehot encoded}
\State $\bm{A} = \text{ones}(N_{\ell},N_{\ell+1})$ \Comment{$\bm{A}$ initialization (such that the mean starting $\bm{\pi}$ is one)}
\EndIf
\State store = []
\For{t in 1:T}  
\For{i in 1:$N_c$}  
\State $\bm{x} \sim \mathcal{N}(\bm{\mu}$[i], $\bm{\sigma}^2$[i]) \Comment{sample lower level data} 
\State $\bm{\pi}$ = $\bm{Ar}$[i] \Comment{Compute confidence estimate}
\State $\bm{\delta} = 0.5(\bm{1}/\bm{\pi}-(\bm{x}-\bm{\mu}$[i])$^2$) \Comment{compute second-order errors}
\State $\bm{A} \leftarrow \bm{A} + \eta \bm{A} \circ \bm{\delta r}[i]^T$ \Comment{update $\bm{A}$ following Eq.~5}
\EndFor
\State store[t] = sum(norm([$\bm{\sigma}^2$[i] - $\bm{1}/\bm{Ar}$[i] for i in 1:$N_c$]))/($\sqrt{N_{\ell}}N_c)$ \Comment{distance between (real) $\bm{\sigma}^2$ and $\bm{1}/\bm{\pi}$}
\EndFor
\end{algorithmic}
\end{breakablealgorithm}

\section{Simulation details: Approximate Bayes-optimal integration}

\begin{figure}[!ht]
\centering
\includegraphics{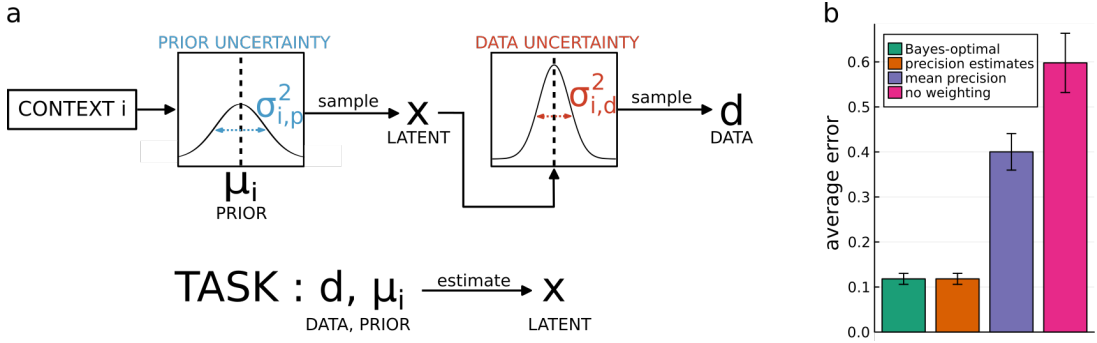}
\caption{Approximate Bayes-optimal computation in a volatile environment. (a) Simulation setup. (b) Simulation results for data and prior integration when the distributiosn of both data and prior are context-dependent.
Bayes-optimal (green): a Bayes-optimal estimate, with knowledge of true prior uncertainty and true data confidence. Precision estimates (orange): our dynamics, with knowledge
of true prior uncertainty and an estimate of data confidence as a function of current representation. Mean confidence (purple): an estimate with knowledge only of the mean prior
uncertainty and data confidence across contexts. No weighting (magenta): an estimate blind to uncertainty and confidence. We plot the average distance between each estimate
and the true latent. The error bars indicate the standard deviation.}
        \label{fig:sibayes}
\end{figure}

For simulations presented in Fig.~\ref{fig:sibayes}b, we follow the simulation procedure described below. Pseudocode for these simulations is presented in Supplementary Algorithm \ref{alg:bayes} and a mathematical intuition is given in S\ref{SI:bayes}a.

We consider a higher area with $N_{\ell+1}$ neurons and a lower area with $N_{\ell}$ neurons. We consider $N_c$ different classes of inputs, each with its own distribution $\mathcal{N}(\bm{\mu_i}, \bm{\sigma_{i}}^2), i \in [1, N_c]$, where $\bm{\mu_i}$ and $\bm{\sigma_{i}}^2$ are vectors of size $N_{\ell}$.  We initialize all $\bm{\mu_i}$ following a $\mathcal{U}(0,2/N_{\ell})$ and all $\bm{\sigma_{i}}^2$ by randomly choosing each component in $\{0.1, 2\}$ with a $50\%$ chance. We initialize the confidence estimation matrix $\bm{A}$ following a $\mathcal{U}(0,2)$. We additionally collect the mean prior variance vector across classes $\bm{\bar{\sigma}}^2 = \frac{1}{N_c}\sum_i \bm{\sigma_i}^2$ and the mean data confidence vector across classes $\bm{\bar{\pi}} = \frac{1}{N_c}\sum_i \bm{A}\phi(\bm{\mu_i})$. We then repeat across epochs the following procedure. For each class $i$ (1) we sample a true target latent $\bm{x} \sim \mathcal{N}(\bm{\mu_i}, \bm{\sigma_{i}}^2)$. We consider that the confidence estimation weights are correct such that the confidence of the data is $\bm{\pi} = \bm{A}\phi(\bm{x})$. (2) Then we sample noisy data. Here we want to focus on confidence estimation and not mean prediction, so we suppose that the prediction function is the identity, and we then sample data $\bm{d} \sim \mathcal{N}(\bm{x}, 1/\bm{\pi})$. The goal is then to infer $\bm{x}$ from data $\bm{d}$ and prior $\bm{\mu_i}$. We do that in four different ways that differ in how they take into account uncertainty and confidence:\\
(3i) a Bayes-optimal estimate, with knowledge of true prior variance and true data confidence 
\begin{equation}
    \bm{u} = (\bm{\pi} \circ \bm{d} + \bm{\sigma_{i}}^2 \circ \bm{\mu_i})/(\bm{\pi}+\bm{\sigma_{i}}^{-2})\;.
\end{equation}
(3ii) our dynamics, with knowledge of true prior variance and data confidence estimation
\begin{equation}
    \tau\dot{\bm{u}} = -\bm{u} + \bm{\mu_i} + \bm{\sigma_{i}}^2 \circ \bm{A}\phi(\bm{u})\circ(\bm{d}-\bm{u})\;.
\end{equation}
(3iii) an estimate with knowledge only of the mean prior variance and data confidence across classes
\begin{equation}
    \tau\dot{\bm{u}} = -\bm{u} + \bm{\mu_i} + \bm{\bar{\sigma}} \circ \bm{\bar{\pi}}\circ(\bm{d}-\bm{u})\;.
\end{equation}
(3iv) an estimate blind to variance and confidence
\begin{equation}
    \tau\dot{\bm{u}} = -\bm{u} + \bm{\mu_i} + (\bm{d}-\bm{u})\;.
\end{equation}
In Fig.~\ref{fig:sibayes}b, we plot the average distance between each estimate and the true latent $(N_cN_e\sqrt{N_{\ell}})^{-1}\|x-u\|$ and its standard deviation. \\

\begin{breakablealgorithm}
\caption{Approximate Bayes-optimal integration}\label{alg:bayes}
\begin{algorithmic}
\Require $N_{\ell+1}$, $N_{\ell}$, $N_c$ $\phi$, $\tau$, T, $N_e$
\State $\bm{\sigma}^2 = [\text{choice}([0.1, 2]) \text{ for \_ in } 1:N_{l}] \text{ for \_ in 1:}N_c]$   \Comment{Initialize prior variance} 
\State $\bm{\mu} = [2\text{rand}(N_{\ell})/N_{\ell} \text{ for \_ in 1:}N_c]$ \Comment{Initialize prior mean} 
\State $\bm{A} = 2\text{rand}(N_{\ell}, N_{\ell+1})$  \Comment{Initialize confidence estimation weights} 
\State $\bm{\bar{\sigma}} = \text{sum}(\bm{\sigma^2})/N_c$ \Comment{mean prior variance} 
\State $\bm{\bar{\pi}} = \text{sum}([\bm{A}\bm{\mu}[i] \text{ for i in } 1:N_c])/N_c$ \Comment{mean data confidence} 
\State err1s, err2s, err3s, err4s = [~], [~], [~], [~]
\For{t in 1:$N_e$}
\For{i in 1 $N_c$}
\State $\bm{x} \sim \mathcal{N}(\bm{\mu}[i], \bm{\sigma}^2[i])$ \Comment{sample true data} 
\State $\bm{\pi} =\bm{A}\phi(\bm{x})$ \Comment{compute confidence estimate at true data} 
\State $\bm{d} \sim \mathcal{N}(\bm{x}, 1/\bm{\pi})$ \Comment{sample noisy data} 
\State $\bm{\hat{x}} = (\bm{\pi}\circ\bm{d}+\bm{\sigma}^{-2}[i]\circ\bm{\mu}[i])/(\bm{\pi}+\bm{\sigma}^{-2}[i])$ \Comment{Bayes-optimal estimate} 
\State err1s.append($\text{norm}(\bm{x}-\bm{\hat{x}})/\sqrt{N_{\ell}}$)
\State $\bm{u} = \bm{1}$
\For{t in 1:T}
    \State $\bm{u} \mathrel{+}= (1/\tau) * (-\bm{u} + \bm{\mu}[i] +  \bm{\sigma}^2[i] \circ \bm{A}\phi(\bm{u}) \circ (\bm{d} - \bm{u}))$ \Comment{dynamics with confidence estimation} 
\EndFor
\State err2s.append($\text{norm}(\bm{x}-\bm{u})/\sqrt{N_{\ell}}$)
\State $\bm{u} = \bm{1}$
\For{t in 1:T}
    \State $\bm{u} \mathrel{+}= (1/\tau) * (-\bm{u} + \bm{\mu}[i] +  \bm{\bar{\sigma}} \circ \bm{\bar{\pi}} \circ (\bm{d} - \bm{u}))$ \Comment{dynamics with average confidence and prior variance} 
\EndFor
\State err3s.append($\text{norm}(\bm{x}-\bm{u})/\sqrt{N_{\ell}}$)
\State $\bm{u} = \bm{1}$
\For{t in 1:T}
    \State $\bm{u} \mathrel{+}= (1/\tau) * (-\bm{u} + \bm{\mu}[i] + (\bm{d} - \bm{u}))$ \Comment{no weighting} 
\EndFor
\State err4s.append($\text{norm}(\bm{x}-\bm{u})/\sqrt{N_{\ell}}$)
\EndFor
\EndFor
\end{algorithmic}
\end{breakablealgorithm}

\section{Simulation details: Nonlinear binary classification}

For simulations presented in Fig.~3 , we built the datasets by sampling $N=1000$ points $(x_1, y_1), \dots, (x_N, y_N)$ from each of the Gaussian distributions represented in Fig.~3di (first column: $\mathcal{N}([0,0], \text{diag}([3,3]))$ and $\mathcal{N}([0,0], \text{diag}([1/3,1/3]))$, second column: $\mathcal{N}([0,0], \text{diag}([1,1/4]))$ and $\mathcal{N}([0,0], \text{diag}([1/4,1/4]))$, third column: $\mathcal{N}([1,0], \text{diag}([1/5,1/5]))$, $\mathcal{N}([-1,0],\\
\text{diag}([1/5,1/5]))$, $\mathcal{N}([0,1], \text{diag}([1/5,1/5]))$ and $\mathcal{N}([0,-1], \text{diag}([1/5,1/5]))$, represented by their 99.7\% confidence ellipses) and attaching the corresponding class label (either red or blue). 

We then build a 2x2 network where the top level activity is a one-hot representation of the class and the bottom level activity is the coordinate in space $(x,y)$.
We train this network in supervised learning settings on the dataset by clamping both top and bottom area to the corresponding elements of the dataset and perform one step of parameters learning as described in Eqs.~4 and 5.

We then test the capacity of our network to classify data by only clamping the bottom level to the data and letting the top level activity follow Eq.~2.
We then select as the output class index the index of the maximum top level activity, and plot the corresponding classification in Fig.~4dii. 

For comparison, we also plot (in Fig.~4diii) the classification results obtained with the same 2x2 architecture but using classical predictive coding dynamics
\begin{equation}
    \tau\dot{\bm{u_{\ell}}} = -\bm{u_{\ell}} + \bm{W_{\ell}r_{\ell+1}} + \bm{r_{\ell}}' \circ \bm{W_{\ell-1}^T}\bm{e_{\ell-1}}\;,
\end{equation}
\begin{equation}
    \dot{\bm{W_{\ell}}} \propto \bm{e_{\ell}}\bm{r_{\ell+1}}^T
\end{equation}
and following the same training and testing procedures.

In Fig.~4e we plot the associated performance, with the addition of the maximum likelihood estimate with perfect knowledge of the means and variances.\\

\begin{breakablealgorithm}
\caption{Training}\label{alg:train}
\begin{algorithmic}
\Require dataset, $\bm{W}$, $\bm{A}$, $\phi$, $\eta_w$, $\eta_a$
\Ensure terms of A are strictly positive, range of $\phi$ is positive
\For{$(\bm{d},\bm{t})$ in dataset}  \Comment{$(\bm{d},\bm{t})$ is ( [x,y] data, one-hot target)}
\State $\bm{\pi} = \bm{A}\phi(\bm{t})$ \Comment{confidence estimate}
\State $\bm{e} = \bm{d} - \bm{W}\phi(\bm{t})$ \Comment{raw error}
\State $\bm{\delta} = 0.5(\bm{1}/\bm{\pi} - \bm{e}^2)$ \Comment{second-order error}
\State $\bm{W} \leftarrow \bm{W} + \eta_w(\bm{\pi} \circ \bm{e})\bm{t}^T$ \Comment{prediction weight learning, Eq.~4}
\State $\bm{A} \leftarrow \bm{A} + \eta_a \bm{A} \circ\bm{\delta}\bm{t}^T$ \Comment{confidence estimation weight learning, Eq.~5}
\EndFor
\end{algorithmic}
\end{breakablealgorithm}

\begin{breakablealgorithm}
\caption{Testing}\label{alg:test}
\begin{algorithmic}
\Require data, $\bm{W}$, $\bm{A}$, $\phi$, $\phi'$ $\tau$, T
\State inferred\_labels = dict()
\State $\bm{t}$ = [0.5, 0.5] \Comment{Uniform initialization of top level}
\For{$\bm{d}$ in data} 
\For{$i=1..\text{T}$}
\State $\bm{\pi} = \bm{A}\phi(\bm{t})$ \Comment{confidence estimate}
\State $\bm{e} = \bm{d} - \bm{W}\phi(\bm{t})$ \Comment{raw error}
\State $\bm{\delta} = 0.5(\bm{1}/\bm{\pi} - \bm{e}^2)$ \Comment{second-order error}
\State $\bm{a} = \phi'(\bm{t}) \circ (\bm{W}^T(\bm{\pi} \circ \bm{e}) + \bm{A}^T\bm{\delta})$ \Comment{Total propagated error Eq.~3}
\State $\bm{t} \leftarrow \bm{t} + \tau^{-1}(-\bm{t} + \bm{a})$ \Comment{Neuronal dynamics Eq.~2 without top down influence}
\EndFor
\State inferred\_labels[$\bm{d}$] = \text{argmax}($\bm{t}$) \Comment{Most probable class index}
\EndFor
\end{algorithmic}
\end{breakablealgorithm}

\end{document}